\documentclass[aps,twocolumn,amsmath,amssymb,preprintnumbers]{revtex4}
\usepackage{amsmath} \usepackage{amsfonts} \usepackage{amssymb}
\usepackage{bbm}
\usepackage{epsfig}
\usepackage{graphics}
\usepackage{graphicx}
\newcommand{\be}{\begin{equation}}
\newcommand{\ee}{\end{equation}}
\newcommand{\ba}{\begin{eqnarray}}
\newcommand{\ea}{\end{eqnarray}}
\newcommand{\nn}{\nonumber}
\newcommand{\kr}{\rangle}
\newcommand{\kl}{\langle}

\newcommand{\hk}{{^k}}

\begin{document}

\title[ ]{Universality of geometry}

\author{C. Wetterich}
\affiliation{Institut  f\"ur Theoretische Physik\\
Universit\"at Heidelberg\\
Philosophenweg 16, D-69120 Heidelberg}

\begin{abstract}

In models of emergent gravity the metric arises as the expectation value of some collective field. Usually, many different collective fields with appropriate tensor properties are candidates for a metric. Which collective field describes the ``physical geometry''? We resolve this ``metric ambiguity'' by an investigation of the most general form of the quantum effective action for several metrics. In the long-distance limit the physical metric is universal and accounts for a massless graviton. Other degrees of freedom contained in the various metric candidates describe very massive scalars and symmetric second rank tensors. They only play a role at microscopic distances, typically around the Planck length. The universality of geometry at long distances extends to the vierbein and the connection. On the other hand, for distances and time intervals of Planck size geometry looses its universal meaning. Time is born with the big bang. 
\end{abstract}

\maketitle

Lattice spinor gravity \cite{LSG}, \cite{CWCS} is a proposal for a regularized theory of quantum gravity based on a Grassmann functional integral for ``fundamental'' fermions. The metric arises as the expectation value of a collective field, typically formed from four (or more) fermions. Similarly, the vierbein can be realized as the expectation value of a fermion bilinear. However, the model contains many different collective fields with appropriate transformation properties of a metric or a vierbein. This poses the problem of the ``metric ambiguity''. Which metric should be chosen for a description of the ``physical geometry''?

For an example with two complex Dirac-spinors $\psi^{\tilde a}, \tilde a=1,2$ we could choose for the vierbein $e^m_\mu$ either
\be\label{A1}
e^{1m}_\mu\sim \kl \psi^{\tilde a} C_1\gamma^m\partial_\mu\psi^{\tilde a}\kr,
\ee
or 
\be\label{A2}
e^{2m}_\mu\sim\kl \psi^1 C_2\gamma^m\partial_\mu\psi^2\kr,
\ee
with $\gamma^m$ Dirac matrices and $C_{1,2}$ appropriate charge conjugation matrices \cite{CWCS}. We also could employ $(\bar\psi=\psi^\dagger\gamma^0)$
\be\label{A3}
e^{3m}_\mu\sim\kl\bar\psi^{\tilde a}\gamma^m\partial_\mu\psi^{\tilde a}\kr.
\ee
Metric candidates are, for some suitable $a$,
\be\label{A4}
g_{1,\mu\nu}=Re(e^{am}_\mu e^{an}_\nu\eta_{mn}),
\ee
or 
\be\label{A5}
g_{2,\mu\nu}\sim Re(\kl\partial_\mu H^+_k\partial_\nu H^-_k+\partial_\nu H^+_k
\partial_\mu H^-_k\kr),
\ee
with $H^\pm_k$ fermion bilinears that are invariant under generalized Lorentz transformations \cite{LSG,CWCS,CWTS}. A priori, the geometries constructed from these various objects differ from each other and it is not clear which one should be chosen. The metric ambiguity (or, more generally, ``geometry ambiguity'') extends to other formulations of spinor gravity \cite{HCW,CWSG} or other models where the vierbein is described by the expectation value of a fermion bilinear \cite{Aka,Ama,Den,Di,GV,FC}. 

The formulation of quantum gravity in terms of fundamental fermions is not crucial for this aspect. A lattice model for quantum gravity based on a bosonic non-linear $\sigma$-model \cite{LDI} shows the same ambiguity. Even proposals for a formulation quantum gravity in terms of fundamental geometrical objects \cite{LQG1,LQG2,LQG3} exhibit the metric ambiguity: besides some metric that may be related directly to objects of lattice geometry in a suitable continuum limit, other fields transforming as symmetric tensors with respect to general coordinate transformations exist as well. 

The metric ambiguity is present in classical gravity as well. Instead of a given metric field $g_{\mu\nu}$ one could also consider an alternative metric $g_{2,\mu\nu}=\alpha g_{\mu\nu}+\beta R_{\mu\nu}+\gamma Rg_{\mu\nu}$ with $R_{\mu\nu}$ the Ricci curvature tensor and curvature scalar $R=R_{\mu\nu}g^{\mu\nu}$. A priori, it is not obvious why distances should not be measured with $g_{2,\mu\nu}$ instead of $g_{\mu\nu}$. Also the formulation of quantum field theory in a given curved space, with background metric $g_{\mu\nu}$, reveals the same problem. Observables of the type 
$\tilde g_{\mu\nu}\sim\partial_\mu(\bar\psi\psi)\partial_\nu(\bar\psi\psi)$ typically acquire a non-zero expectation value. (In the language of the effective action this is due to a term linear in $\bar g_{\mu\nu}=\kl \tilde g_{\mu\nu}\kr$ of the type $\int_x\sqrt{g}g^{\mu\nu}\bar g_{\mu\nu}$.) Again, one could consider $g_{2,\mu\nu}=\alpha g_{\mu\nu}+\beta\bar g_{\mu\nu}$ as the ``physical metric''. 

On a deeper level, distances and geometry can be extracted from correlation functions \cite{CWG}. The metric obtains from derivatives of suitably normalized connected two-point functions. The ambiguity concerns now the choice of the correlation function which is used for the definition of a distance.

In this note we propose a very simple general solution to the problem of the metric ambiguity. In the long distance limit it simply does not matter which one of the various metric candidates one chooses for measuring distances and defining the geometry. The different metrics are simply proportional to each other, up to tiny corrections. The distances measured with anyone of the possible metrics are the same, up to a choice of units. The same holds for the vierbein. Geometry is universal.

Our discussion is based on the most general form of the effective action for the metric. This effective action (generating functional of $1$PI-correlation functions) includes the quantum fluctuations and specifies the field equations. The symmetries of general coordinate transformations (and local Lorentz-transformations in the presence of fermions) permit a mixing of different metric candidates. As a result, only one independent metric remains in the long distance limit. This can be associated with the ``physical metric'' and it describes the graviton. 

The other degrees of freedom contained in the fields for the various metric candidates describe supermassive scalar and tensor particles. Typical masses are of the order of the Planck mass. In the ``universal limit'' for wavelengths much larger than the Planck length all effects from the exchange of those heavy fields become tiny and can be neglected. It is this decoupling of the heavy modes which makes geometry universal. Of course, the existence of the universal long wavelength limit requires that nontrivial gravitational physics exists at scales large compared to the Planck length. As usual for gravity, this requires a vanishing cosmological constant $\lambda$ - or, more precisely, $|\lambda|\ll M^4$, with $M$ the reduced Planck mass.

\medskip\noindent
\textit{\textbf{Two metric fields}}

The key points for a resolution of the metric ambiguity can be understood within a simple model for two metric fields $g_{1,\mu\nu}$ and $g_{2,\mu\nu}$. We assume that the quantum effective action for these two fields is invariant under general coordinate transformations. (With respect to diffeomorphisms both fields are symmetric covariant tensors. 
Infinitesimal transformations obey $\delta g_{a,\mu\nu}=-\xi^\rho\partial_\rho g_{a,\mu\nu} - \partial_\mu\xi^\rho g_{a,\rho\nu}-\partial_\nu\xi^\rho g_{a,\mu\rho},a=1,2$.) The field equations are obtained from the variation of the effective action. We concentrate first on the part of the effective action which does not involve derivatives. This ``metric potential'' $V(g_{1,\mu\nu},g_{2,\mu\nu})$ is the only relevant piece for static and homogeneous solutions. 

If only one metric is present the metric potential is dictated by diffeomorphism symmetry to take the form
\be\label{1}
V=\lambda\sqrt{g}\ ,\ g=|\det g_{\mu\nu}|.
\ee
Static and homogeneous solutions of the field equations have to obey
\be\label{2}
\frac{\partial V}{\partial g_{\mu\nu}}=\frac{1}{2} \lambda\sqrt{g}\, g^{\mu\nu}=0\ ,\ g^{\mu\nu}g_{\nu\rho}=\delta^\mu_\rho.
\ee
Regular solutions require a fine-tuning $\lambda=0$ - this is the usual issue of a vanishing cosmological constant. For vanishing $\lambda$ the metric is not determined by the potential. The derivative terms, i.e. the Einstein-Hilbert action involving the curvature scalar, typically admit flat space solutions $g_{\mu\nu}=c\eta_{\mu\nu}$. (We keep the signature of $\eta_{\mu\nu}$ arbitrary. The constant $c$ can be set to unity by a rescaling of coordinates.)

For two metric fields the general form of the metric potential becomes more involved. For regular $g_{1,\mu\nu},g_{2,\mu\nu}$ we can now use both $\sqrt{g_1}$ and $\sqrt{g_2}$. Furthermore, scalars can be constructed as $g_{1,\mu\nu}g^{\mu\nu}_2$ - this is not possible for a single metric field since by definition of the inverse metric one has $g_{a,\mu\nu} g^{\nu\rho}_a=\delta^\rho_\mu$ for every $a$. The crucial aspects can be seen by considering a metric potential of the form
\ba\label{3}
V&=& \alpha_1\sqrt{g_1}+\alpha_2\sqrt{g_2}\nonumber\\
&&+(\beta_1\sqrt{g_1}+\beta_2\sqrt{g_2})g_{1,\mu\nu}g_2^{\mu\nu}\nonumber\\
&&+(\delta_1\sqrt{g_1}+\delta_2\sqrt{g_2})g_{2,\mu\nu}g_1^{\mu\nu}.
\ea
Higher order terms, involving for example $g_{1,\mu\nu}g_{1,\rho\sigma}g_2^{\mu\rho}g_2^{\nu\sigma}$, do not affect the qualitative features.

Consider first the ansatz
\be\label{4}
g_{1,\mu\nu}=\sigma_1\eta_{\mu\nu}\ ,\ g_{2,\mu\nu}=\sigma_2\eta_{\mu\nu}.
\ee
For nonzero and finite $\sigma_a$ we use the ratio $\gamma=\sigma_2/\sigma_1$  such that the metric potential reads
\be\label{5}
V=\sigma^2_1\left(\sum_n s_n\gamma^n\right)=\sigma^2_1 W(\gamma).
\ee
Here $n$ takes values in the range $[-1,3]$, with $s_{-1}=4\beta_1,\ s_0=\alpha_1,\ s_1=4(\beta_2+\delta_1),\ s_2=\alpha_2,\ s_3=4\delta_2$. Extending the potential \eqref{3} by inclusion of terms quartic in $g_{\mu\nu}$ or $g^{\mu\nu}$ adds terms with $n=-2$ and $n=4$ and changes the coefficients $s_n$. The form $V=\sigma^2_1 W(\gamma)$ is preserved for an arbitrary metric potential. This is a direct consequence of diffeomorphism symmetry, since  besides $\sqrt{g}$ every term in the potential must involve an equal number of metric fields and inverse metric fields. (Beyond the expansion \eqref{5} $W(\gamma)$ can become an arbitrary function.)

A solution with ansatz \eqref{4} requires an extremum of $W(\gamma)$ at some $\gamma_0$, with $W(\gamma_0)=W_0$.
Generically, such an extremum exists. Insertion of this partial extremum yields $V=W_0\sigma^2_1$, and we recognize the potential \eqref{1} with $\lambda=W_0$. A static and homogeneous solution requires again a tuning of parameters such that $W_0=0$. A possible solution is then  $g_{1,\mu\nu}=\eta_{\mu\nu},\ g_{2,\mu\nu}=\gamma_0\eta_{\mu\nu}$. As an example we choose $\alpha_1=\alpha_2=-1,\ \beta_1=\beta_2=\delta_1=\delta_2=\frac{1}{8}$ such that  $\gamma_0=1, \ W(\gamma_0)=0$.

\medskip\noindent
\textit{\textbf{Stability of Minkowski space}}

We next expand the metric potential around $g_{2,\mu\nu}=\gamma g_{1,\mu\nu}$ with
\be\label{6}
g_{1,\mu\nu}=g_{\mu\nu}\ ,\ g_{2,\mu\nu}=\gamma (\delta^\rho_\mu+f_\mu{^\rho})g_{\rho\nu}.
\ee
Using the determination of $\gamma$ by $\partial W/\partial\gamma=0$ the terms linear in $f_\mu{^\nu}$ vanish. (More generally, the requirement of a vanishing linear term $\sim f=f_\mu{^\nu} \delta^\mu_\nu$ defines the value of $\gamma$.)
In quadratic order one obtains
\be\label{7}
V=\sqrt{g}\left\{\lambda+\frac{1}{2}\mu f^2+\frac{1}{4}\nu\tilde f_\mu{^\nu}\tilde f_\nu{^\mu}\right\},
\ee
with
\ba\label{8}
f=\tilde f_\mu{^\nu}\delta^\mu_\nu~,~
\tilde f_\mu{^\nu}=f_\mu{^\nu}-\frac14 f\delta^\nu_\mu~,~\tilde f_\mu{^\nu}\delta^\mu_\nu=0,
\ea
and
\ba\label{9}
\lambda&=&\alpha_1+\alpha_2\gamma^2+\frac{4}{\gamma}(\beta_1+\beta_2\gamma^2)+4\gamma(\delta_1+\delta_2\gamma^2),\nonumber\\
\mu&=&\frac{\beta_1}{2\gamma}+\frac{\alpha_2}{8}\gamma^2+\frac{3\delta_2}{2} \gamma^3,\nonumber\\
\nu&=&4\frac{\beta_1}{\gamma}-\alpha_2\gamma^2-4\delta_2\gamma^3.
\ea
(For our example one has $\lambda=0,\ \mu=\frac{1}{8},\ \nu=1$.)  This result points to a straightforward interpretation: The model describes a standard gravitational theory with metric $g_{\mu\nu}$. In addition, there is a massive scalar field $f$ with squared mass $\sim \mu$,  and a massive traceless tensor field $\tilde f_{\mu}{^\nu}$ with squared mass $\sim \nu$. A model with two metrics amounts then to a model with one metric plus additional massive fields. If one tunes $\lambda=0$ only one field, the graviton $g_{\mu\nu}$, is massless. This picture extends to the most general form of the metric potential $V$ provided the parameters are such that $W(\gamma)$ has an extremum, $\mu$ and $\nu$ are positive, and $\lambda=0$ is possible by appropriate tuning.

For a discussion of the stability of the flat space  solution we need the kinetic terms in the effective action for $g_{1,\mu\nu}$ and $g_{2,\mu\nu}$. They can be written in terms of $g_{\mu\nu},\ f$ and $\tilde f_{\mu}{^\nu}$. We assume first that for
long-wavelength modes  a derivative expansion becomes valid. Then a possible diffeomorphism invariant kinetic term with up to two derivatives reads
\ba\label{10}
{\cal L}_{\rm kin}&=&\sqrt{g}\left\{-\frac{M^2}{2}R+\frac{1}{2}Z\partial^\mu f \partial_\mu f\right.\nn\\
&&+\frac{1}{4}W_1
D^\rho\tilde f_\mu{^\nu} D_\rho \tilde f_\nu{^\mu}
+\frac{1}{2} W_2
D_\rho\tilde f_\mu{^\rho} D_\sigma\tilde f_\nu{^\sigma} g^{\mu\nu}\nn\\
&&+\frac{1}{2}W_3\partial^\nu fD_\mu\tilde f_\nu{^\mu}\Big\}.
\ea
(Here indices are raised and lowered with $g^{\mu\nu}$ and $g_{\mu\nu}$, while the covariant derivative $D_\mu$ and the curvature scalar $R$ are formed with $g_{\mu\nu}$.) Neglecting  for a moment $\tilde f_{\mu}{^\nu}$ this yields for a Minkowski signature $\eta_{\mu\nu}={\rm diag} (-1,1,1,1)$ a standard theory of a free massive scalar field coupled to gravity.  Stability of Minkowski space is ensured for $M^2>0,\ Z>0$ by the positive energy theorem. The mass of the scalar field is given by $m=(\mu/Z)^{\frac{1}{2}}$. If no special choice of parameters occurs the generic size of the couplings is $\mu,\nu,\lambda\sim M^4,\ Z,W_k\sim M^2$ with $M$ the reduced Planck mass. The scalar field is therefore very massive, with $m\sim M$. 

The general discussion of stability  for the tensor modes $\tilde f_{\mu\nu}$ is more involved. For our purpose it is  sufficient that a region in the space of couplings $W_k$ exists for which linear stability of Minkowski space is realized. (For example this holds for $W_1>0,\ |W_k|\ll W_1$. A discussion of the stability issue can be found in ref. \cite{RW}.) 

More general diffeomorphism invariant kinetic terms turn $M^2,Z$ and $W_k$ into functions of the invariants $f,\tilde f_\mu{^\nu}\tilde f_\nu{^\mu}~,~\tilde f_\mu{^\nu}\tilde f_\nu{^\rho}\tilde f_\rho{^\mu}$ etc.. This induces a mixing between different modes. However, a function $M^2 f_\mu{^\nu}$ can be brought to a constant $M^2$ by a suitable Weyl scaling of $g_{\mu\nu}$, thereby modifying $Z$ and $W_{k}$. After Weyl scaling we may evaluate the couplings $Z,W_{k}$ as the values of the corresponding functions at $f_{\mu}{^\nu}=0$. An expansion of $Z,W_k$ around these values accounts for interactions. In this sense eq. \eqref{10} can be interpreted as the most general diffeomorphism invariant kinetic term with two derivatives and up to two powers of $f_{\mu}{^\nu}$. 

It is instructive to discuss the linearized theory for the case $W_2=W_3=0$. The linearized field equations
\be\label{15A}
(W_1\partial^2-\nu)\tilde f_\mu{^\nu}
\ee
do not mix the different components of $\tilde f_\mu{^\nu}$. For positive $\nu$ and $W_1$ there is no tachyon, such that for Minkowski space the wave solutions for $\tilde f_\mu{^\nu}$ are stable. The nine components of $\tilde f_\mu{^\nu}$ describe states with spin two (five components obeying the constraint $D_\nu\tilde f_\mu{^\nu}=0$), spin one (three components) and spin zero (one component). For $W_2=W_3=0$ the corresponding particles all have an equal mass $\sqrt{\nu/W_1}$. This degeneracy is lifted for $W_{2,3}\neq 0$. (Neither $W_2$ nor $W_3$ contribute to the spin two mode. For $W_3\neq 0$ also the spin zero states in $f$ and $\tilde f_\mu{^\nu}$ are mixed.) By continuity, the squared mass of all particles stays positive for not too large values of $W_2$ and $W_3$ such that stability of Minkowski space with respect to linear fluctuations is maintained.

Beyond the linear level the issue of stability gets more complicated. For $W_2=W_3=0$ we may compute the energy density
\be\label{15B}
\rho=T_{00}=\frac{2}{\sqrt{g}}\frac{\delta\Gamma_f}{\delta g^{00}},
\ee
where the variation with respect to $g^{00}$ is performed for fixed $\tilde f_\mu{^\nu},\mu\leq \nu$, and $f$, while $\Gamma_f$ subtracts from $\Gamma$ the term $\sim R$ which does not depend on the ``matter fields'' $f$ and $\tilde f_\mu{^\nu}$. One finds (recall $\tilde f_0{^0}=-\sum_k\tilde f_k{^k}$)
\ba\label{15C}
\rho&=&\frac12\mu f^2+\frac14\nu\
\Big \{ \sum_k (\tilde f_k{^k})^2 +\Big(\sum_k\tilde f_k\hk\Big)^2\Big\}\nn\\
&&+\frac12\nu\sum_k\Big\{\Big(\tilde f_0{^k}\Big)^2
+\sum_{l>k}(\tilde f_k{^l})^2\Big\}\nn\\
&&+\frac12 Z (\partial_0 f)^2+\frac12 Z\sum_i(\partial_i f)^2\nn\\
&&+\frac14 W_1\Big\{\sum_i\Big[\sum_k(\partial_i\tilde f_k\hk)^2+(\partial_i\sum_k\tilde f_k\hk)^2\nn\\
&&+2\sum_{k,l>k}(\partial_i\tilde f_k{^l})^2+2\sum_k(\partial_i\tilde f_0\hk)^2\Big]\nn\\
&&+\sum_k(\partial_0\tilde f_k\hk)^2+(\partial_0\sum_k\tilde f_k\hk)^2\nn\\
&&+2\sum_{k,l>k}(\partial_0\tilde f_k{^l})^2
-6\sum_k(\partial_0\tilde f_0\hk)^2\Big\}.
\ea
(We have indicated explicitly here the sum over the space indices $k,l,i=1\dots 3$.) For obtaining eq. \eqref{15C} one has to take into account that for symmetric $g_{2,\mu\nu}$ the relation
\be\label{15D}
\tilde f_\mu{^\nu}=g_{\mu\rho}g^{\nu\sigma}\tilde f_\sigma{^\rho}
\ee
involves the metric, such that 
\ba\label{15E}
\frac{\delta}{\delta g^{00}}(\tilde f_\mu{^\nu}\tilde f_\nu{^\mu})&=&2\sum_k(\tilde f_0\hk)^2,\nn\\
\frac{\delta}{\delta g^{00}}(D^\rho\tilde f_\mu{^\nu} D_\rho\tilde f_\nu{^\mu})&=&
\partial_0\tilde f_\mu{^\nu}\partial_0\tilde f_\nu{^\mu}-2\sum_k(\partial_0\tilde f_0\hk)^2\nn\\
&&+2\sum_k\sum_i(\partial_i\tilde f_0\hk)^2.
\ea
All terms in $\rho$ are positive except the last one. Similarly the contribution of the term $\sim W_2$ to the energy density reads
\ba\label{15F}
\rho_2=\frac12 W_2\{\partial_i\tilde f_i\hk \partial_j\tilde f_j\hk+
(\partial_0\tilde f_0{^0}+\partial_i\tilde f_0{^i})^2\nn\\
+2\partial_0\tilde f_0\hk\partial_i\tilde f_i\hk-3\partial_0\tilde f_0\hk\partial_0\tilde f_0\hk\}.
\ea

For $W_2=W_3=0$ the energy density can become negative for the lower spin modes. At this stage we can therefore not invoke the positive energy theorem in order to guarantee stability of Minkowski space on the non-linear level. It is not clear if for the action \eqref{7}, \eqref{10} stability can be ensured by other conserved quantities or by a positive energy for a suitable range of parameters with non-vanishing $W_{2,3}$. If not, stability may also be realized by a non-local form of the effective action \cite{CWNL}. Such non-local terms can arise from a local formulation with additional fields by integrating out the additional fields. A similar situation is familiar for the treatment of massive spin one particles. For our purpose we will be satisfied here with linear stability. 

At this point we can generalize our setting to an arbitrary number $n$ of metrics $g_{a,\mu\nu},\ a=1\ldots n$. The effective action describes a standard graviton $g_{\mu\nu}=g_{1,\mu\nu}$  plus $n-1$ scalars and $n-1$ traceless symmetric tensors contained in $f_{a,\mu\nu}=f_{a,\mu}{^\rho}g_{\rho\nu}=g_{a,\mu\nu}/\gamma_a-g_{\mu\nu},\ a=2\ldots n$. Minkowski space is stable for a large range of effective couplings (which are generic up to the tuning $\lambda=0$). The setting resembles in many aspects the situation for higher dimensional theories which lead after dimensional reduction to a massless graviton and an (infinite) tower of massive ``Kaluza-Klein-gravitons''.

\medskip\noindent
\textit{\textbf{Universal gravity}}

At this point we can answer the question ``what is the physical metric?''. In fact, for the long-distance behavior it does not matter which metric  we choose. We could equivalently define $g_{\mu\nu}=g_{2,\mu\nu}$ and make an expansion for $g_{1,\mu\nu},\ g_{1,\mu\nu}=(1/\gamma)(g_{2,\mu\nu}+f'_{\mu\nu})$. The massive scalar and tensor modes are now described by $f'_{\mu\nu}=f'_\mu{^\rho}g_{2,\rho\nu}$. As long as the massive modes can be neglected the metrics  $g_{1,\mu\nu}$ and $g_{2,\mu\nu}$ are simply proportional to each other, such that the distances measured with the two metrics are the same up to an overall proportionality constant. The difference between the use of $g_{1,\mu\nu}$ or $g_{2,\mu\nu}$ for defining the ``physical metric'' reduces to a choice of units. It disappears if we measure length in units of the Planck length. One can always rescale the metric such that $M^2$ in eq.\eqref{10} has a given value.

Differences in the geometry determined by $g_{1,\mu\nu}$ or $g_{2,\mu\nu}$ appear only for non-vanishing $f_{\mu}{^\nu}$. With couplings of $g_{1,\mu\nu}$ and $g_{2,\mu\nu}$ to other particles (e.g. ``matter'' or ``radiation'') of a similar size  we can compute the relative size of such differences. Indeed for small momenta $q_\mu,\ q^2=q^\mu q_\mu$,  the response of the metric to a source $T$ reads  $h\sim q^{-2} M^{-2} T$. (We work here in momentum space and omit indices for $h_{\mu\nu}=g_{\mu\nu}-\eta_{\mu\nu}$ and  the energy momentum tensor $T_{\mu\nu}$.) On the other hand, the response of $f$ obeys $f\sim (q^2+M^2)^{-2}M^{-2} T$. In position space and for a static source this results in an exponential suppression $f\sim\exp(-Mr)h$. Thus  $f$ becomes negligible for distances large compared to the Planck length. More generally, for all processes with characteristic energies and momenta small compared to $M$ the differences between the geometries measured by $g_{1,\mu\nu}$ or $g_{2,\mu\nu}$ become tiny. 

Only for a particular tuning of parameters the mass of one of the tensors may be small. These special cases correspond to bimetric theories \cite{E2}. Models for one light and one massless tensor \cite{F2} do, in general, not show a universal geometry. For this type of models a universal geometry requires additional assumptions for the couplings of the two metric fields to matter and radiation. Modifications of late cosmology and a possible role for dark energy \cite{G2} are only expected for the special cases where one of the tensors is light. In this note we stick to the generic case where the additional tensor fields are superheavy and long-distance gravity as well as late cosmology are universal. 

\medskip\noindent
\textit{\textbf{Euclidean instability}}

Let us consider a setting where the signature of the metric is not fixed a priori. For $\lambda=0$ flat space is a solution of the field equations derived from the effective action $\Gamma=\int_x(V+{\cal L}_{\rm{kin}})$ for an arbitrary signature of $\eta_{\mu\nu}$. In particular, both Minkowski space and euclidean space are a solution. The stability properties of these two solutions are different, however. For appropriate signs of $\mu,\nu, M^2,Z$, and with $W_k$ in an appropriate range, Minkowski space is stable with respect to small deviations. These deviations describe propagating tensor and scalar waves corresponding to the graviton and massive scalars and traceless tensors. 

The situation for euclidean space is different. We may try to associate one of the coordinates with time $t$ and consider solutions of the field equation which are specified by initial conditions at $t=0$. Then the possible solutions of a field equation of the type
\be\label{11}
\partial^2\varphi=(\partial^2_t+\partial^2_x)\varphi=0
\ee
typically show a strong growth for increasing time. (For simplicity we consider only one space coordinate here.) For example, a small local deviation from flat space with initial conditions
\be\label{12}
\varphi(x,t=0)=\varphi_0\exp \{-x^2/a^2\}~,~\partial_t\varphi(x,t=0)=0,
\ee
grows as
\be\label{13}
\varphi(x,t)=\varphi_0\exp\left(-\frac{x^2}{a^2}\right)\cos \left(\frac{2xt}{a^2}\right)\exp\left(\frac{t^2}{a^2}\right).
\ee
For large enough $t$ the field values for $\varphi$ exceed any bound. This contrasts with Minkowski space, where $(\partial^2_t-\partial^2_x)\varphi=0$ leads for the same initial conditions \eqref{12} to the wave solution
\be\label{14}
\varphi(x,t)=\frac{\varphi_0}{2}\left[\exp\left\{-\frac{(x+t)^2}{a^2}\right\}+\exp
\left\{-\frac{(x-t)^2}{a^2}\right\}\right],
\ee
which remains bounded.

Furthermore, flat space is not a local minimum of the euclidean action. (See ref. \cite{CWNL} for a detailed discussion.) For positive $M^2$ the physical scalar mode 
\ba\label{15}
\zeta&=&h-\frac43 \frac{\partial^\mu\partial^\nu}{\partial^2}\tilde h_{\mu\nu},\nn\\
h&=&h_{\mu\nu}\delta^{\mu\nu}~,~\tilde h_{\mu\nu}=h_{\mu\nu}-\frac14 h\delta_{\mu\nu},
\ea
has a negative kinetic term 
\be\label{16}
{\cal L}_{\rm kin}(\zeta)=-\frac{3M^2}{64}\partial^\mu\zeta\partial_\mu\zeta,
\ee
while the kinetic term for the traceless transverse tensor is positive. Flat space is a saddlepoint of the euclidean action, not a minimum. In contrast to Minkowski space, no positive energy theorem forbids the growth of small deviations from flat space. 

Many euclidean quantum field theories have a unique bounded solution of the field equation, with a ``vacuum state'' or ``ground state'' that is homogeneous in all directions (and therefore static.) One may ask if the saddle point behavior of flat space implies that a flat euclidean geometry is not a possible stable vacuum solution. This conclusion too strong, however, since the saddle point behavior follows from the assumption of a derivative expansion \eqref{10} for small gradients. Flat space can be a possible vacuum, but the effective action has to involve non-local terms in this case, as discussed in \cite{CWNL}. If euclidean flat space is realized as a vacuum, all correlation functions will decay for large time differences. In contrast to Minkowski space, no ``long-time memory'' is possible. 

\medskip\noindent
\textit{\textbf{Vierbein and connection}}

In the presence of fermions the formulation of general relativity and geometry should be based on the vierbein $e^m_\mu$ and a spin connection $\omega_{\mu mn}$ \cite{Ca}. In standard gravity the metric obtains from the vierbein as $g_{\mu\nu}=e^m_\mu e^n_\nu\eta_{mn}$. The spin connection, as well as the Levi-Civita connection, is computable in terms of the vierbein. If we do not want to fix the signature of the metric a priori we may work with a complex vierbein. We can then employ $\eta_{mn}=\delta_{mn}$, while Minkowski space is realized for $e^k_\mu=\delta^k_\mu~,~k=1,2,3$, and $e^0_0=i$.

Different geometries can now be realized by the use of different vierbeins. The resolution of this ambiguity in the long wavelength limit proceeds in analogy to the metric ambiguity. Consider two vierbeins $e^{1m}_\mu$ and $e^{2m}_\mu$. The ``vierbein potential'' can now be constructed from powers of $e^{am}_\mu$ and their inverse $e^{a\mu}_m$, with $e^{am}_\mu e^{a\nu}_m=\delta^\nu_\mu$ (no sum over $a$). In particular, expressions of the type $(\alpha_1\sqrt{g_1}+\alpha_2\sqrt{g_2})$ in eq. \eqref{3} are generalized to 
\be\label{17}
e=\frac{1}{24}\epsilon^{\mu_1\dots \mu_4}\epsilon_{m_1\dots m_4}
A_{a_1\dots a_4}e^{a_1m_1}_{\mu_1}\dots e^{a_4m_4}_{\mu_4}.
\ee
Here $A$ is totally symmetric in the four ``flavor indices'' $a_1\dots a_4$. With respect to general coordinate transformations $e$ is a scalar density, similar to $\sqrt{g}$. (In standard gravity, with only one vierbein, one has $e=\det (e^m_\mu)=\sqrt{g}.)$

We can again make the ansatz
\be\label{18}
e^{1m}_\mu=e^m_\mu~,~e^{2m}_\mu=\tilde\gamma e^m_\mu
\ee
and obtain for the most general form of the vierbein potential
\be\label{19}
V=\det(e^m_\mu)\tilde W(\tilde \gamma).
\ee
Homogeneous solutions require an extremum of $\tilde W(\tilde\gamma)$ at $\tilde\gamma_0$, with $\tilde W(\tilde\gamma_0)=0$. (Now $\tilde W(\tilde\gamma_0)$ plays the role of the cosmological constant $\lambda$.) The generalization to an arbitrary number of vierbeins, with a corresponding extended range for the flavor index $a$, is straightforward. Including suitable diffeomorphism and (generalized) Lorentz invariant kinetic terms for $e^{am}_\mu$ the spectrum of excitations around flat space (for $\tilde W(\tilde\gamma_0)=0)$ comprises the massless vierbein $e^m_\mu$ as well as massive tensors contained in $e^{am}_\mu-\tilde\gamma^a e^m_\mu,a\geq 2$. For suitable parameters Minkowski space is stable, while the issue of euclidean instability remains present in the vierbein formulation. 

What about the relation between the metric and the vierbein? A priori, possible candidates for a metric do not have to obey the relation $g_{\mu\nu}=e^m_\mu e^n_\nu\eta_{mn}$. The issue can be cast into the ``two-metric formalism''. We may identify $e^m_\mu e^n_\nu\eta_{mn}=g_{1,\mu\nu}$, and some other possible metric candidate with $g_{2,\mu\nu}$. In the long-distance limit only one metric, say $g_{\mu\nu}=g_{1,\mu\nu}$, survives, while $g_{2,\mu\nu}-\gamma e^m_\mu e^n_\nu\eta_{mn}$ describes heavy tensor modes. The setting is the same as for generalized gravity discussed in ref. \cite{CWDRG}. The relation $g_{\mu\nu}=e^m_\mu e^n_\nu\eta_{mn}$ arises as a universal relation for long wavelengths, while on a microscopic scale deviations from this relation are expected due to the role of the heavy tensor modes. (For a complex vierbein one may use $g_{\mu\nu}=Re(e^m_\mu e^n_\nu\delta_{mn})$. Then $Im(e^m_\mu e^n_\nu\delta_{mn})$ is associated to heavy tensor modes.) 

This discussion extends to the role of connections. From the vierbein $e^m_\mu$ and its first derivatives one can construct the spin connection $\omega(e)_{\mu mn}$ and the Levi-Civita connection $\Gamma_{\mu\nu}{^\lambda}$. (The Levi-Civita connection can be expressed in terms of the metric $g_{\mu\nu}=e^m_\mu e^n_\nu\eta_{mn}$ and its first derivative.) Consider now any other candidate $\tilde \omega_{\mu mn}$ for a connection. The difference between two connections is a tensor 
\be\label{20}
\tilde\omega_{\mu mn}=\omega(e)_{\mu mn}+K_{\mu mn}.
\ee
A general effective action involves a heavy mass for the tensor $K_{\mu mn}$. In the long distance limit one therefore obtains the universal relation $\tilde\omega_{\mu mn}=\omega(e)_{\mu mn}$, with the associated Levi-Civita connection $\Gamma_{\mu\nu}{^\lambda}$. The relations between the various connections and tensors are described in more detail in the context of generalized gravity \cite{CWDRG}. 

\medskip\noindent
\textit{\textbf{Birth of time}}

In conclusion, we have discussed a simple and very general mechanism that leads to a universal geometry in situations where more than one candidate for a metric is present. It relies on the effective decoupling of heavy modes. Universality of geometry is then realized on length scales that are large compared to the range of interactions mediated by the heavy modes. In momentum space, universality occurs for energies or momenta that are small compared to the mass of the heavy modes. 

The other facet of this setting is the loss of universality on distance scales that are close to the inverse mass of the heavy particles. This issue touches directly fundamental questions of the type: What happened before the big bang? Is there a a beginning of time or is time infinite? These questions assume the existence of a metric that can be used to measure time intervals. If there is no universal metric, there is no universal time. Thus the answer to the question of what happens with time close to the Planck time may simply be: ``There is no unique time anymore.''

For a given fundamental theory many collective fields exist whose expectation value can serve as a metric. In the same spirit, many different correlation functions can be used to define a distance and a geometry. Assume now that such a model has a characteristic scale $m$ - either due to the presence of couplings with dimension of length or mass, or to dimensional transmutation from running dimensionless couplings, or else generated by spontaneous breaking of (approximate) dilatation symmetry. This scale $m$ will be the natural mass scale for the heavy tensor modes discussed in this note. It seems natural (although not strictly necessary) to assume that the Planck mass $M$ is in the vicinity of $m$. The fate of time in the Planck era is then the loss of its universal meaning. Once many different metrics can be used on equal footing for a measurement of time intervals there is no more time in the usual sense of a universal quantity. In the evolution of the universe a universal time emerges only once characteristic distances in space and time exceed $m^{-1}$. In this sense, time is born with the big bang.

\end{document}